\documentclass[%
reprint,
twocolumn,
amsmath,amssymb,
aps,
prb,
superscriptaddress 
]{revtex4-2}

\usepackage{bm}
\usepackage{times}
\usepackage{amsmath}
\usepackage{mathrsfs}
\usepackage{graphicx}
\usepackage{epstopdf}
\usepackage[colorlinks=true, letterpaper=ture, pdfstartview=FitV, linkcolor=blue, citecolor=blue, urlcolor=blue]{hyperref}
\usepackage{dcolumn}

\begin{document}

\title{Magnon-magnon interaction in monolayer MnBi$_2$Te$_4$}

\author{Yiqun Liu}
\affiliation{National Laboratory of Solid State Microstructures and School of Physics, Nanjing University, Nanjing 210093, China}
\affiliation{Collaborative Innovation Center of Advanced Microstructures, Nanjing University, Nanjing 210093, China}

\author{Liangjun Zhai}
\thanks{E-mail: zhailiangjun@nju.edu.cn}
\affiliation{National Laboratory of Solid State Microstructures and School of Physics, Nanjing University, Nanjing 210093, China}
\affiliation{Collaborative Innovation Center of Advanced Microstructures, Nanjing University, Nanjing 210093, China}
\affiliation{School of Mathematics and Physics, Jiangsu University of Technology, Changzhou 213001, China.}

\author{Songsong Yan}
\affiliation{National Laboratory of Solid State Microstructures and School of Physics, Nanjing University, Nanjing 210093, China}
\affiliation{Collaborative Innovation Center of Advanced Microstructures, Nanjing University, Nanjing 210093, China}

\author{Di Wang}
\affiliation{National Laboratory of Solid State Microstructures and School of Physics, Nanjing University, Nanjing 210093, China}
\affiliation{Collaborative Innovation Center of Advanced Microstructures, Nanjing University, Nanjing 210093, China}

\author{Xiangang Wan}
\thanks{E-mail: xgwan@nju.edu.cn}
\affiliation{National Laboratory of Solid State Microstructures and School of Physics, Nanjing University, Nanjing 210093, China}
\affiliation{Collaborative Innovation Center of Advanced Microstructures, Nanjing University, Nanjing 210093, China}

\date{\today}

\begin{abstract}

MnBi$_2$Te$_4$, the first confirmed intrinsic antiferromagnetic topological insulator, has garnered increasing attention in recent years.
Here we investigate the energy correction and lifetime of magnons in MnBi$_2$Te$_4$ caused by magnon-magnon interaction. 
First, a calculation based on the density functional theory was performed to get the parameters of the magnetic Hamiltonian of MnBi$_2$Te$_4$. 
Subsequently, the perturbation method of many-body Green's function was employed and the first-order self-energy [$\Sigma^{(1)}(\bm k)$] and second-order self-energy [$\Sigma^{(2)}(\bm k,\varepsilon_{\bm k})$] of magnon were obtained. 
Numerical computations reveal that the corrections from both $\Sigma^{(1)}(\bm k)$ and $\Sigma^{(2)}(\bm k,\varepsilon_{\bm k})$ strongly rely on momentum and temperature, with the energy renormalization near the Brillouin zone (BZ) boundary being significantly more pronounced than that near the BZ center.
Furthermore, our findings indicate the occurrence of dip structures in the renormalized magnon spectrum near the $\rm K$ and $\rm M$ points. These dip structures are determined to be attributed to the influence of $\Sigma^{(2)}(\bm k,\varepsilon_{\bm k})$.

\end{abstract}
\maketitle

\section{Introduction}

Recently, there has been a growing interest in two-dimensional (2D) insulating magnets, both in fundamental and applied research domains \cite{1-McGuire,1-Dai,1-Mak,1-Lin,1-GongS}. 
Among the various 2D insulating materials, MnBi$_2$Te$_4$ (MBT) stands out as the pioneering intrinsic antiferromagnetic (AFM) topological insulator (TI) \cite{1-Gong,1-Li22,1-Zhang,1-Otrokov}.
This material offers a promising platform for exploring unique topological phenomena, including the quantum anomalous Hall (QAH) effect \cite{1-Otrokov2,1-Deng}, topological axion state \cite{1-Liu,1-Chen}, and Weyl semimetal state \cite{1-Zhang,1-Li22}.
MBT is composed of Te-Bi-Te-Mn-Te-Bi-Te septuple layers (SL) arranged along the $c$ axis, resulting in an A-type antiferromagnetic (AFM) ground state, and crystallizes in a rhombohedral structure with the space group $R\bar3m$\cite{1-Ning,1-Yan}. 
Apart from its topological properties, the magnetism of MBT is also of great significance \cite{1-Riberolles,1-Chen1,1-Ge}.

Magnons, which are quanta of spin waves, serve as a fundamental concept in characterizing the magnetic properties of materials.
The behavior of magnons in MBT, as an antiferromagnetic insulator, has also captured the interest of researchers.
Experimental studies, such as inelastic neutron scattering (INS), have provided valuable insights into the properties of MBT. These studies have revealed the Ising-like nature and magnetic frustration present in MBT. Furthermore, they have shown that the magnon spectrum in MBT exhibits a significant broadening of lifetime \cite{1-Lii}. 
In the realm of atomically thin MBT layers, Raman spectroscopy has revealed distinct characteristic peaks corresponding to magnon excitations. This observation serves as compelling evidence for the existence of long-range magnetic order in the thin-layer limit of MBT \cite{1-Lujan}.
A theoretical study employing a self-consistent renormalized (SCR) spin-wave theory has discovered that interactions between magnons can lead to notable renormalization of the magnon spectrum in MBT \cite{1-Wei}. This renormalization effect is found to be dependent on the momentum, particularly in the vicinity of the high-symmetry points.

In the magnon representation of a spin Hamiltonian, magnon-magnon interactions are typically treated as small terms, often expanded using techniques such as Holstein-Primakoff \cite{1-Holstein,1-Oguchi} or Dyson-Maleev transformations \cite{1-Dyson1,1-Dyson2,1-Maleev}.
The terms in the Hamiltonian involving more than two magnon operators correspond to the interaction terms, specifically known as the multiple magnon processes.
As the temperature increases, the influence of magnon-magnon interactions becomes more pronounced, resulting in significant effects such as apparent linewidth broadening \cite{1-Harris} and energy renormalization of magnons.
In addition to the aforementioned effects, multiple magnon processes that do not conserve the number of magnons can even lead to spontaneous decay of magnons at zero temperature \cite{1-Canali,1-Igarashi,1-Zhitomirsky}. 
These scattering processes can introduce intriguing momentum or temperature-dependent behaviors of magnons in MBT, which have not been previously explored in the study conducted using the SCR spin-wave theory \cite{1-Wei}.

In present work, the magnetic model of MBT is constructed by considering exchange and anisotropy terms. The exchange integrals are determined through a density functional theory (DFT) calculation. Subsequently, the perturbation methods of many-body Green's function are employed to derive the first-order and second-order self-energies of magnons. 
To avoid encountering a negative second-order self-energy around the $\Gamma$ point, a long-wavelength approximation of the momentum of thermal magnons is implemented on the second-order self-energy \cite{1-Pershoguba}. A systematic investigation is carried out to explore the impact of multiple magnon processes on the spectrum and lifetime of magnons in MBT.

\section{Model Hamiltonian and Formalism}\label{Sec-2}
\subsection{The magnon Hamiltonian of MBT}
\begin{figure}[t]
\centering
\includegraphics[scale=1.2]{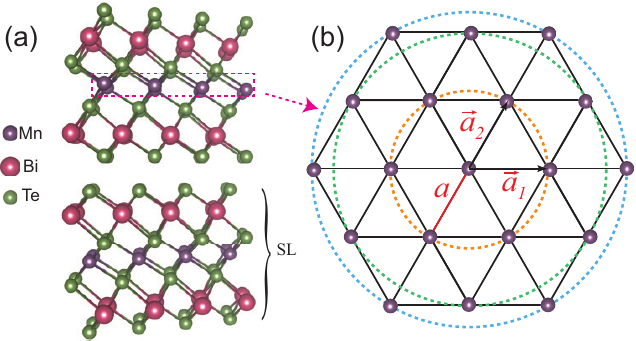}
\caption{(a) Lattice structure of MBT; two Te-Bi-Te-Mn-Te-Bi-Te layers are shown here. (b) A schematic view of the Mn layer in an SL, where $\vec a_1, \vec a_2$ denote the lattice vector and $a$ denotes the lattice constant.}\label{Gra-Lattice}
\end{figure}

The MBT crystal is composed of seven atomic blocks arranged in the sequence of Te-Bi-Te-Mn-Te-Bi-Te, forming a stacked structure along the $c$ axis in an ABC-type arrangement \cite{1-Gong,1-Lee4}, as depicted in Fig. \ref{Gra-Lattice}(a). 
In a single SL, the Mn atoms form a 2D triangle lattice, as depicted in Fig. \ref{Gra-Lattice} (b).
As a result of the magnetic frustration in MBT, the INS experiment indicates that  \cite{1-Lujan}, in addition to the exchange interaction amongst the nearest neighbors, the longer-range exchange interaction and the single-ion anisotropy play crucial roles in stabilizing the magnetic structure within a single SL.
Hence the magnetic Hamiltonian describing a single SL of MBT can be expressed as follows:
\begin{equation}\label{HH}\begin{aligned}
H=&-J_1\sum_{\langle i,j\rangle}S_i\cdot S_j-J_2\sum_{\langle\langle i,j\rangle\rangle}S_i\cdot S_j\\
&-J_3\sum_{\langle\langle\langle i,j\rangle\rangle\rangle}S_i\cdot S_j-K\sum_i\left(S_i^z\right)^2,
\end{aligned}\end{equation}
where $\langle i,j\rangle$,  $\langle\langle i,j\rangle\rangle$, and $\langle\langle\langle i,j\rangle\rangle\rangle$ represent the summation over the nearest, next-nearest, and third-nearest neighbor sites, respectively.
$J_1, J_2$ and $J_3$ represent the pairwise interactions between the nearest neighboring,  the next-nearest neighboring, and the third-nearest neighboring ions within a triangular layer, respectively. Additionally, $K$ refers to the single-ion anisotropy.

To obtain the magnon Hamiltonian, the magnetic Hamiltonian is usually transformed using the Holstein-Primakoff transformation. This transformation is preferred as it ensures the preservation of the hermiticity of the Hamiltonian \cite{1-Oguchi,1-White}. The Holstein-Primakoff transformation can be expressed as follows \cite{1-Holstein}:
\begin{equation}\label{EqHP}
\begin{aligned}
&S_i^+=\sqrt{2S}\left(1-\frac 1{4S}a_i^\dagger a_i\right)a_i,\\
&S_i^-=\sqrt{2S}a_i^\dagger\left(1-\frac 1{4S}a_i^\dagger a_i\right),\\
&S_i^z=S-a_i^\dagger a_i,
\end{aligned}
\end{equation}
where $a_i^\dagger$ and $a_i$ represent the creation and annihilation operators for magnons, respectively.
By substituting Eq.~(\ref{EqHP}) into Eq.~(\ref{HH}) and retaining terms up to fourth order in the boson creation and annihilation operators, the magnon Hamiltonian beyond the linear spin-wave theory is obtained.
The Fourier transformation of magnon operators is introduced as
\begin{equation}
\label{EqFourier}
a_i=\frac1{\sqrt N}\sum_ka_k e^{i{\bm k}\cdot{\bm r}_i},\quad a_i^\dagger=\frac1{\sqrt N}\sum_{\bm k}a_{\bm k}^\dagger e^{-i{\bm k}{\bm r}_i}.
\end{equation}
After performing the Fourier transformation, the magnon Hamiltonian in $\bm k$ space can be expressed as
\begin{eqnarray}\label{HK}
H&=&\sum_{\bm k}\varepsilon_{\bm k}a_{\bm k}^\dagger a_{\bm k},\\ \nonumber
&&+\frac1N\sum_{\bm1,\bm2,\bm3,\bm4}\delta(\bm1+\bm2-\bm3-\bm4)V_{\bm1\bm2\bm3\bm4}a_{\bm1}^\dagger a_{\bm2}^\dagger a_{\bm3}a_{\bm4},
\end{eqnarray}
where the subscripts $\bm1, \bm2, \bm3, \bm4$ correspond to the momenta $\bm k_1$, $\bm k_2$, $\bm k_3$, $\bm k_4$, respectively.
$\varepsilon_{\bm k}$ represents the bare magnon dispersion, which can be expressed as
\begin{equation}
\varepsilon_{\bm k}=\sum_{l=1}^3Z_lJ_lS(1-\gamma_{\bm k}^{(l)})+K(2S-1),
\end{equation}
where the coordination numbers are given by $Z_1=Z_2=Z_3=6$.
The structure factor is defined as $Z_l\gamma_k^{(l)}=\sum_{\bm\delta_l}e^{i\bm k\cdot\bm\delta_l}$, where $\bm\delta_l$ ($l=1,2,3$) represents the vector to the $l$th neighbor.

The second term in Eq.~(\ref{HK}) describes the interaction between magnons.
The vertex function $V_{\bm1\bm2\bm3\bm4}$ quantifies the strength of the magnon-magnon interaction in Eq.~(\ref{HK}), and it can be expressed as
\begin{equation}\begin{aligned}
V_{\bm1\bm2\bm3\bm4}=\sum_{l=1}^3\frac{Z_lJ_l}8\big(\gamma_{\bm1}^{(l)}&+\gamma_{\bm2}^{(l)}+\gamma_{\bm3}^{(l)}+\gamma_{\bm4}^{(l)}\\
&-2\gamma_{\bm2-\bm3}^{(l)}-2\gamma_{\bm2-\bm4}^{(l)}\big) -K.
\end{aligned}\end{equation}

\subsection{Green's function}
In this section, we will introduce the perturbation methods of the many-body Green's function \cite{1-Mahan}.
The magnon Matsubara function is defined as follows:
\begin{equation}
\mathcal G({\bm k},\tau-\tau')=-\langle\mathcal T_\tau a_{\bm k}(\tau)a_{\bm k}^\dagger(\tau')\rangle,
\end{equation}
The bare magnon Matsubara function is expressed as $\mathcal G^0({\bm k},i\omega_n)=1/(i\omega_n-\varepsilon_{\bm k})$.

\begin{figure}[htpb]
\centering
\includegraphics[scale=1.0]{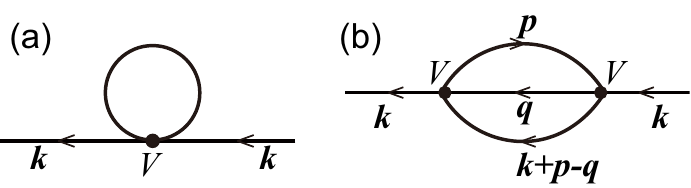}
\caption{Feynman diagrams corresponding to the (a) first-order and (b) second-order self-energies of magnons; the solid line represents a magnon and the arrow denotes the propagation direction.}\label{Gra-2}
\end{figure}

To investigate the impact of magnon-magnon interaction, the diagram expansion method of the Green's function is employed \cite{1-Mahan}.
The first-order and second-order self-energy diagrams resulting from magnon-magnon interactions are depicted in Fig. \ref{Gra-2}.  Notably, the second-order diagram represents the lowest order diagram that contributes to the decay rate of magnons \cite{1-Harris}.
Based on the first-order Hartree diagram shown in Fig. \ref{Gra-2}(a), the resulting self-energy $\Sigma^{(1)}(\bm k)$ can be expressed as follows: 
\begin{equation}\label{Eq-6}
\Sigma^{(1)}(\bm k)=-\sum_{l=1}^3Z_lJ_lS(1-\gamma_{\bm k}^{(l)})\big[\beta(T)-\eta_l(T)\big]-4KS\beta(T),
\end{equation}
where the temperature-dependent parameters are defined as
$\beta(T)=\frac1{NS}\sum_{\bm p}n_{\bm p},\,\eta_l(T)=\frac1{NS}\sum_{\bm p}\gamma_{\bm p}^{(l)}n_{\bm p}$
, and $n_{\bm p}=\langle a_{\bm p}^\dagger a_{\bm p}\rangle$ represents the Bose distribution function.
$\Sigma^{(1)}(\bm k)$ does not contribute to magnon decay, but it does cause a temperature and momentum-dependent renormalization of the magnon energy $\varepsilon_k$. This effect has been thoroughly discussed in previous literature \cite{1-Rezende}.

The diagram shown in Fig. \ref{Gra-2}(b) represents the second-order diagram.
From this diagram, the second-order self-energy $\Sigma^{(2)}({\bm k},i\omega)$ is derived as
\begin{eqnarray}\label{Eq-ZiNeng-p-q}
\Sigma^{(2)}({\bm k},i\omega_n)=&&-\frac8{N^2}\sum_{\bm p,\bm q}\big|V_{\bm k,\bm p,\bm q,\bm k+\bm p-\bm q}\big|^2\\ \nonumber
                              && \times \dfrac{n_{\bm p}\big[1+n_{\bm q}+n_{\bm k+\bm p-\bm q}\big]-n_{\bm q}n_{\bm k+\bm p-\bm q}}{i\omega_n+\varepsilon_{\bm p}-\varepsilon_{\bm q}-\varepsilon_{\bm k+\bm p-\bm q}}.
\end{eqnarray}
An analytic continuation should be performed on the Matsubara frequency: $i\omega_n\,\rightarrow\,\omega+i\delta$.
As demonstrated in Eq. (\ref{Eq-ZiNeng-p-q}), it is evident that $\Sigma^{(2)}({\bm k},\omega)$ possesses an imaginary component, thereby leading to the decay of magnons.

In the context of the scattering process $\varepsilon_{\bm k}+\varepsilon_{\bm p}\rightarrow\varepsilon_{\bm q}+\varepsilon_{\bm k+\bm p-\bm q}$,
it is crucial to note that the incident energy $\omega$ in $\Sigma^{(2)}({\bm k},\omega)$ should be equal to the energy of the magnon under consideration $\varepsilon_{\bm k}$.
Therefore, considering the contributions of $\Sigma^{(1)}(\bm k)$ and $\Sigma^{(2)}({\bm k},\varepsilon_{\bm k})$,
the renormalized magnon energy can be calculated by
\begin{equation}\label{Eq-ek-all}
\tilde \varepsilon_{\bm k}=\varepsilon_{\bm k}+\Sigma^{(1)}(\bm k)+{\rm Re}\Sigma^{(2)}(\bm k,\varepsilon_{\bm k}).
\end{equation}

\subsection{Long wave-length approximation of thermal magnons}
When studying the interaction between magnons, calculating Eq. (\ref{Eq-ZiNeng-p-q}) directly often poses challenges \cite{1-Igarashi,1-Canali,1-Pershoguba}. 
For instance, when calculating the second-order self energies in an antiferromagnetic system, a common issue that arises is the divergence occurring at the $\Gamma$ point \cite{1-Igarashi,1-Canali}.
In the case of ferromagnetism, direct calculations have revealed that the real part of the second-order self-energy ${\rm Re}\Sigma^{(2)}(\bm k,\varepsilon_{\bm k})$ is consistently negative around the $\Gamma$ point \cite{1-Pershoguba}. 
If only the Heisenberg exchange interaction is considered in the magnetic Hamiltonian (\ref{HH}), both the bare magnon energy $\varepsilon_{\bm k}$ and the first-order self-energy $\Sigma^{(1)}(\bm k)$ in Eq. (\ref{Eq-ek-all}) vanish at the $\Gamma$ point.
When the negative ${\rm Re}\Sigma^{(2)}(\bm k,\varepsilon_{\bm k})$ is included, the renormalized magnon energy $\tilde \varepsilon_{\bm k}$ at the $\Gamma$ point becomes negative. The presence of a negative $\tilde \varepsilon_{\bm k}$ indicates that the magnetic structure is unstable \cite{1-Lujan}.

The presence of negative ${\rm Re}\Sigma^{(2)}(\bm k,\varepsilon_{\bm k})$ can be attributed to the nonvanishing behavior of  $V_{\bm k,\bm p;\bm q,\bm k+\bm p-\bm q}$ as $\bm k$ approaches the $\Gamma$ point. 
To resolve this issue,  a long wavelength approximation with $\bm p\approx0$ is utilized \cite{1-Wang}. After this approximation,  the interaction matrix $V_{\bm k,\bm p=0;\bm q,\bm k+\bm p-\bm q}$ vanishes when $\bm k=0$; thus the problem is resolved.
 This approximation is reasonable when temperature is low, as it assumes that the thermal magnons with momentum $\bm p$ are mostly concentrated near the $\Gamma$ point. The vertex function can be expanded as
\begin{equation}
V_{\bm k,\bm p;\bm q,\bm k+\bm p-\bm q}\approx V_0+\bm p\cdot \bm v_{\bm k;\bm q}+\mathcal O(\bm p^2),
\end{equation}
where $V_0$ represents $\left.V_{\bm k,\bm p;\bm q,\bm k+\bm p-\bm q}\right|_{\bm p=0}$. Considering energy conservation, we obtain $V_0=-K$. The vector function $\bm v_{k;q}$ is given as
\begin{equation}\label{vvv}
\bm v_{\bm k;\bm q}=\left.\Big[\nabla_{\bm p}V-\nabla_{\bm p}E\frac{\nabla_{\bm q}V\cdot\nabla_{\bm q}E}{\left|\nabla_{\bm q}E\right|^2}\Big]\right|_{{\bm p}=0},
\end{equation}
where $V$ indicates $V_{\bm k,\bm p;\bm q,\bm k+\bm p-\bm q}$ and $E=\varepsilon_{\bm k}+\varepsilon_{\bm p}-\varepsilon_{\bm q}-\varepsilon_{\bm k+\bm p-\bm q}$. The second term in Eq. (\ref{vvv}) is obtained by imposing the constraint of energy conservation. The expression for $\bm v_{\bm k;\bm q}$ is given by
\begin{equation}
\bm v_{\bm k;\bm q}=\frac14\sum_{l=1}^3Z_lJ_l\nabla_{\bm p}\left.(\gamma_{\bm k+\bm p-\bm q}^{(l)}-\gamma_{\bm p-\bm q}^{(l)})\right|_{\bm p=0},
\end{equation}
which vanishes as $\bm k=0$. Finally, under the approximation of $\bm p\approx0$, $\Sigma^{(2)}(k,\varepsilon_{\bm k})$ can be expressed as,
\begin{equation}\label{Eq-ZiNeng-q}\begin{aligned}
\Sigma^{(2)}(\bm k,\varepsilon_{\bm k})=&\dfrac{\alpha_1}N\sum_{\bm q}\dfrac{8V_0^2k_BT}{\varepsilon_{\bm k}+\varepsilon_{\bm 0}-\varepsilon_{\bm q}-\varepsilon_{\bm k-\bm q}+i\delta}\\
&+\dfrac{\alpha_2}N\sum_{\bm q}\dfrac{4\big|{\bm v}_{\bm k;\bm q}\big|^2(k_BT)^2}{\varepsilon_{\bm k}+\varepsilon_{\bm 0}-\varepsilon_{\bm q}-\varepsilon_{\bm k-\bm q}+i\delta},
\end{aligned}\end{equation}
where $k_B$ is the Boltzmann constant and
\begin{equation*}
\alpha_1=\frac{\sqrt3}{8\pi}\dfrac{-\ln(1-e^{-C})}{J_{\rm eff}S},\alpha_2=\frac{\sqrt3}{8\pi}\frac{{\rm Li}_2[e^{-C}]}{J_{\rm eff}^2S^2},
\end{equation*}
with $C=K(2S-1)/(k_BT)$, $J_{\it eff}=3J_1/2+9J_2/2+6J_3$, and ${\rm Li}_s(z)$ denoting the polylogarithm \cite{1-Jonquiere}. In this work, the value of $\delta$ in Eq. (\ref{Eq-ZiNeng-q}) is is chosen to be $0.05meV$.

Let us provide a brief discussion on Eq. (\ref{Eq-ZiNeng-q}).
For the magnetic Hamiltonian without the anisotropic terms, i.e., $K$ in Eq. (\ref{HH}) is zero,  it can be observed that $V_0=0$ and $\bm v_{\bm k=0,\bm q}=0$ at the $\Gamma$ point.
As a result, $\Sigma^{(2)}(\bm k,\varepsilon_{\bm k})$ in Eq. (\ref{Eq-ZiNeng-q}) is also zero at the $\Gamma$ point. This ensures that the renormalized magnon energy becomes zero at the $\Gamma$ point \cite{1-Solyom}.
For the magnetic Hamiltonian including anisotropic terms, i.e., when $K$ in Eq. (\ref{HH}) is nonzero, we find that $V_0$ does not vanish at the $\Gamma$ point due to the presence of anisotropy. This anisotropy induces a negative $\Sigma^{(2)}(\bm k,\varepsilon_{\bm k})$ at the $\Gamma$ point.
However, this negative $\Sigma^{(2)}(\bm k,\varepsilon_{\bm k})$ does not lead to a negative renormalized magnon energy. This is because the anisotropy also induces a spin gap at the $\Gamma$ point, which is $(2S)^2$ larger than the negative contribution from ${\rm Re}\Sigma^{(2)}(\bm k,\varepsilon_{\bm k})$ \cite{1-Igarashi}. 
Thus the renormalized magnon energy should still remain positive at the $\Gamma$ point.
It is worth noting that the negative energy contributions from the second-order self-energy can potentially be compensated by higher-order magnon self-energies. However, calculating higher-order magnon self-energies can often be challenging. Therefore, the application of the long-wavelength approximation for thermal magnons is necessary.

Moreover, we find that $\Sigma^{(2)}(\bm k,\varepsilon_{\bm k})$ in Eq. (\ref{Eq-ZiNeng-q}) are divided into two parts.
The first term, which is proportional to $T$, is associated with the anisotropy, while the second term, which is proportional to $T^2$, is related to the isotropic exchange interaction.
Due to the small magnitude of the single-ion anisotropy, the isotropic exchange interaction dominates. As a result, the second-order self-energy still follows the $T^2$ law, which is consistent with the properties of 2D magnetic systems \cite{1-Bayrakci,1-Harris2,1-Bayrakci2}.

\section{Calculation methods}
The calculations, based on the DFT, have been carried out by using the full potential linearized augmented plane-wave method as implemented in the {\it Wien2k} package \cite{1-Blaha}. The $\bm k$-point mesh convergence test has been done (see the Appendix) and a $16\times16\times1$ $\bm k$-point mesh is used for the Brillouin-zone integral. The self-consistent calculations are considered to be converged when the difference in the total energy of the crystal does not exceed $0.00001mRy$. We adopt the generalized gradient approximation of Perdew-Burke-Ernzerhof (PBE-GGA) type \cite{1-Perdew} as the exchange-correlation potential and include the spin orbit coupling (SOC) using the second-order variational procedure \cite{1-Koelling}. We also perform the first-principles calculations with different exchange-correlation potentials and the results are close (see the Appendix). Meanwhile, the GGA + {\it U} scheme is adopted to properly describe the strongly correlated system of Mn 3d orbitals with the effective on-site Coulomb interaction {\it U}$_{\it eff}=2$ $eV$ \cite{1-Dudarev,1-Anisimov}. In order to properly model a monolayer MnBi$_2$Te$_4$ system, i.e., one septuple-layer block, we add a vacuum layer of about $30\mathring{A}$ along the $c$-axis. Minimizing the energy for structural relaxation, we get the lattice parameters as $a=4.43\mathring{A}$, $b=4.43\mathring{A}$, $c=41.18\mathring{A}$. 

The spin exchange interactions are calculated by the open-source software package {\it WienJ} \cite{1-Wang-xx}, which is based on combining magnetic force theorem and linear-response approach \cite{1-Liechtenstein,1-Wan-xx,1-Wan-x1,1-Wang-cc,1-Prange,1-Prange-2}. As an interface to the linearized augmented plane wave (LAPW) software {\it Wien2k} \cite{1-Blaha}, the software allows one to friendly and efficiently calculate spin Hamiltonian parameters. This approach has been successfully applied to various magnetic materials \cite{1-Wang-xx,1-Wan-xx,1-Wan-x1,1-Wan-x2,1-Wan-x3,1-Wang-x2}.

To estimate the Curie temperature $T_c$, Monte Carlo (MC) simulations are performed with the Metropolis algorithm for the Heisenberg model as implemented in {\small MCSOLVER} code \cite{31-Metropolis,32-Liu}, in which a $28\times28\times1$ supercell is adopted. The details of the convergence test for cell size are displayed in the Appendix. The number of MC steps to make the system enter balanced states is 160000, while the number of MC steps involved in measuring is set to 320000.

\section{Results and discussion}\label{Sec-3}

Using the theoretical methods mentioned earlier, we have calculated and obtained the exchange interactions $J_1, J_2, J_3$ in the microscopic magnetic Hamiltonian of MBT, as shown in Table \ref{tab-2},
\begin{table}[h]
\centering
\caption{The values of $J_1,J_2,J_3$ are taken from the INS data and the DFT results, all values are in meV.}\label{tab-2}
\begin{tabular}{c  c c c c c }
\hline
\quad               \;\;& $J_1$        & $J_2$  & $J_3$   \\\hline
Our Results                 \;&\; 0.249 \;&\; -0.024 \;&\; -0.010   \\
INS Data Ref. \cite{1-Liii}    \;&\; 0.233(2) \;&\; -0.033(2) \;&\; 0.007(2)   \\
DFT Data Ref. \cite{1-Li22}   \;&\; 0.25 \;&\; -0.033 \;&\; -  \\
DFT Data Ref. \cite{1-Liiii}   \;&\; 0.22 \;&\; - \;&\; -  \\
\hline
\end{tabular}
\end{table}

The results indicate that $J_1$ is the dominant factor contributing to the ferromagnetism of MBT monolayer, with a value of $0.249 meV$. Furthermore, both $J_2$ and $J_3$ are approximately one-tenth of the magnitude of $J_1$. These results are in good agreement with both INS and DFT results \cite{1-Liii,1-Li22,1-Liiii}. Besides, we adopt the single ion anisotropy $K = 0.048meV$ obtained from the INS data \cite{1-Lii}. Based on these parameters, a MC simulation is performed with the Metropolis algorithm for the Heisenberg model \cite{31-Metropolis,32-Liu}. The obtained Curie temperature is 15.9$K$, which is close to the experimental data 15.2$K$\cite{1-Yang}. 
With the exchange parameters $J_1, J_2, J_3$ and the single ion anisotropy $K$ shown above, we begin the study of the energy spectrum and self-energy of magnons in MBT.

\subsection{Properties of the first-order self-energy}

\begin{figure}[t]
\centering
\includegraphics[scale=0.72]{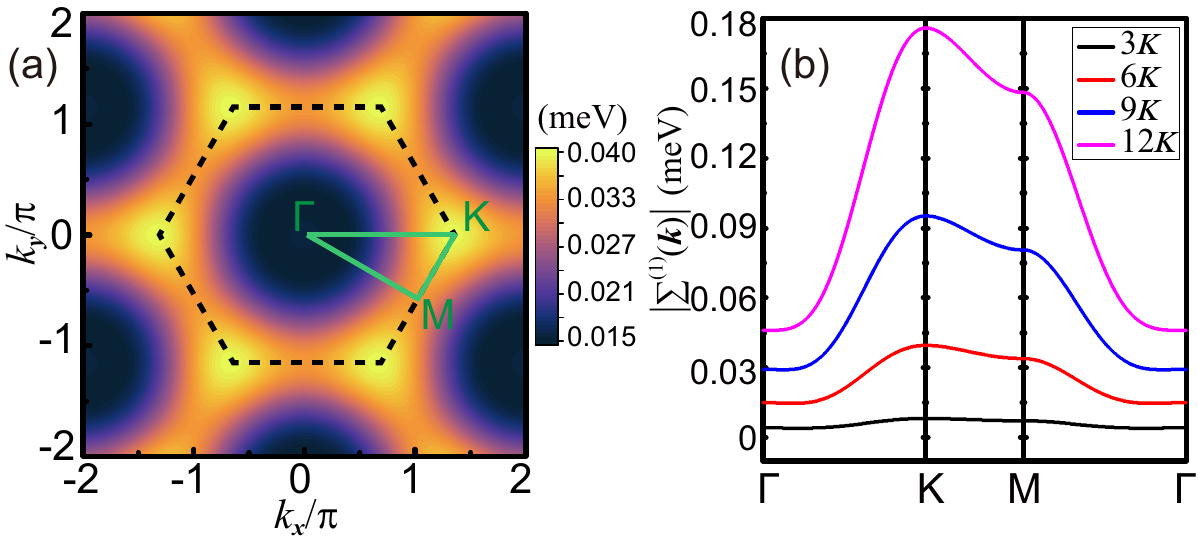}
\caption{(a) Map of $|\Sigma^{(1)}(\bm k)|$ of MBT at $T=6K$ in the $\bm k$ space and (b) $|\Sigma^{(1)}(\bm k)|$ at different temperatures along the high-symmetry line. The black dashed line in (a) represents the boundary of first BZ and the green line is the high-symmetry line along $\rm \Gamma-M-K-\Gamma$. 
}\label{Gra-ZiNeng1-Map}
\end{figure}
First, the properties of first-order self-energy $\Sigma^{(1)}(\bm k)$ are studied.
In Fig. \ref{Gra-ZiNeng1-Map}(a), the magnitude of $\Sigma^{(1)}(\bm k)$ is plotted at a temperature of $6K$ in the $\bm k$ space.
It is shown that the values of $|\Sigma^{(1)}(\bm k)|$ near the boundary of the first BZ are significantly higher compared to those in the central region.
The peak values of $|\Sigma^{(1)}(\bm k)|$ are observed at the high symmetry point $\rm K$, while $\rm M$ serves as the saddle point for $|\Sigma^{(1)}(\bm k)|$.
The behavior of $\Sigma^{(1)}(\bm k)$ indicates that the strength of the magnon-magnon scattering near the boundary of the first Brillouin zone is significantly stronger compared to that in the central region.

In Fig. \ref{Gra-ZiNeng1-Map}(b), the $\bm k$-dependent behavior of $|\Sigma^{(1)}(\bm k)|$ along the high symmetry line $\rm \Gamma-K-M-\Gamma$ is depicted for various temperatures.
Given that the properties of $\Sigma^{(1)}(\bm k)$ are determined by $\beta(T)$ and $\eta_l(T)$, we provide the values of $\beta(T)$ and $\eta_l(T)$ for different temperatures in Table \ref{tab-1}. 
The difference between these values in Table \ref{tab-1} and the results obtained using the SCR spin-wave theory \cite{1-Wei} is not that significant. Here the self-consistent calculation of $\Sigma^{(1)}({\bm k})$ is not done.
It is observed that the values of the parameters increase with increasing temperature, indicating that a greater number of magnons with higher energy can be excited as the temperature rises. Consequently, the scattering probability increases and the renormalization of magnon energy through $\Sigma^{(1)}(\bm k)$ becomes more evident, as illustrated in Fig. \ref{Gra-ZiNeng1-Map}(b).

\begin{table}[h]
\centering
\caption{The values of $\beta,\eta_1,\eta_2,\eta_3$ at different temperatures.}\label{tab-1}
\begin{tabular}{r | c c c c}
\hline
$T$ \;\;& $\beta$ & $\eta_1$  & $\eta_2$  & $\eta_3$   \\  \hline
$3K$    \;&\; 0.008 \;&\; 0.008 \;&\; 0.006 \;&\; 0.006  \\
$6K$    \;&\; 0.031 \;&\; 0.026 \;&\; 0.019 \;&\; 0.016 \\
$9K$   \;&\; 0.061 \;&\; 0.048 \;&\; 0.031 \;&\; 0.026 \\
$12K$   \;&\; 0.096 \;&\; 0.070 \;&\; 0.043 \;&\; 0.035 \\
\hline
\end{tabular}
\end{table}
\begin{figure}[h]
\centering
\includegraphics[scale=0.75]{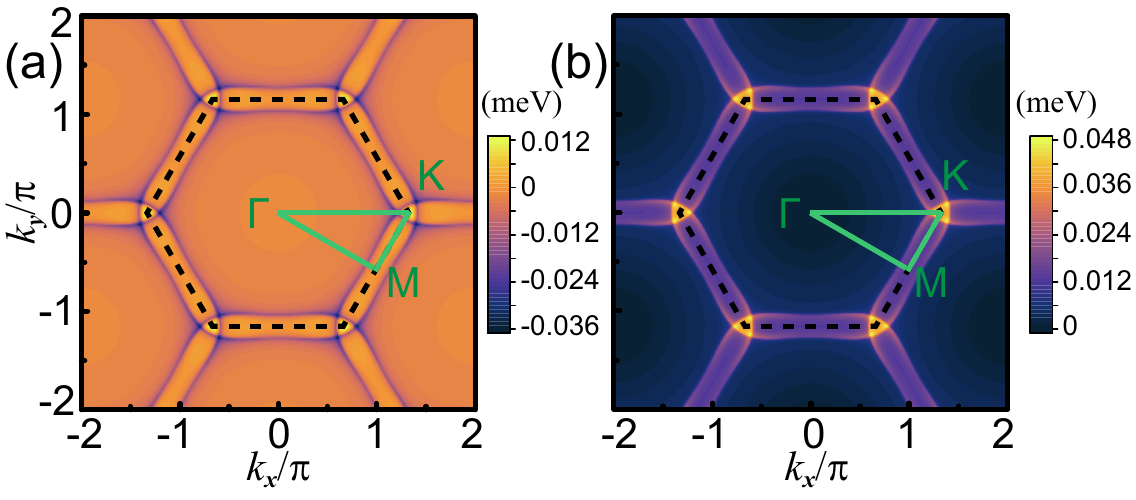}
\caption{Map of (a) ${\rm Re}\Sigma^{(2)}(\bm k,\varepsilon_{\bm k})$ and (b) $\big|{\rm Im}\Sigma^{(2)}(\bm k,\varepsilon_{\bm k})\big|$ of MBT at $T=6K$ in the reciprocal space. The black dashed line represents the boundary of first BZ and the green lines are the high-symmetry lines along $\rm \Gamma-M-K-\Gamma$.
}\label{Gra-ZiNeng-Map}
\end{figure}

\subsection{Second-order self-energy}

In Fig.~\ref{Gra-ZiNeng-Map}, the plot shows the real and imaginary parts of $\Sigma^{(2)}(\bm{k},\varepsilon_{\bm{k}})$ in the $\bm{k}$ space.
In regions distant from the first BZ boundary, both ${\rm Re}\Sigma^{(2)}(\bm{k},\varepsilon_{\bm{k}})$ and $\big|{\rm Im}\Sigma^{(2)}(\bm{k},\varepsilon_{\bm{k}})\big|$ exhibit smooth variations with the wave vector $\bm{k}$. However, as the wave vector $\bm{k}$ approaches the first BZ boundary, these quantities vary sharply.
The maximum values of ${\rm Re}\Sigma^{(2)}(\bm k,\varepsilon_{\bm k})$ and $\big|{\rm Im}\Sigma^{(2)}(\bm k,\varepsilon_{\bm k})\big|$ are observed around the $\rm K$ point.

\begin{figure}[htpb]
\centering
\includegraphics[scale=0.8]{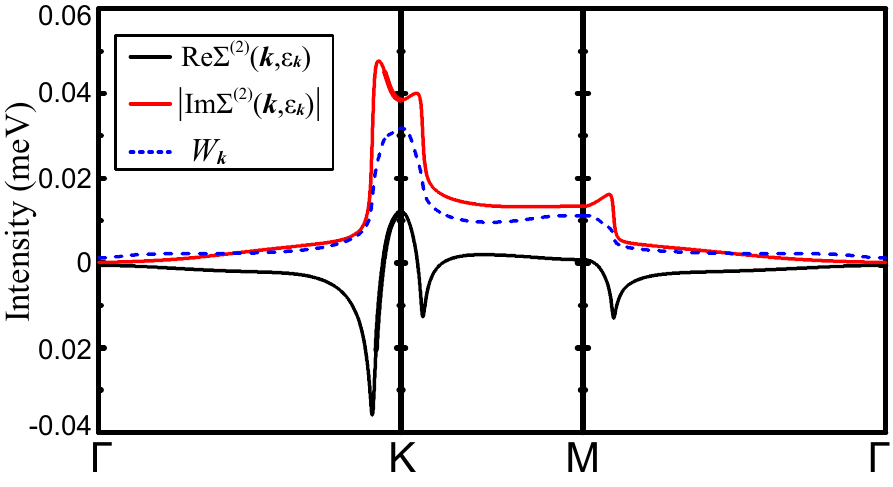}
\caption{Real (black line) and imaginary (red line) parts of the second-order self-energy calculated by Eq. (\ref{Eq-ZiNeng-q}) at $T=6K$ along the $\rm \Gamma-K-M-\Gamma$ line; the blue dashed line represents the momentum dependence of the scattering density of states $W_{\bm k}$. Here we use $\lambda=1.5$.}\label{Gra-ZiNeng}
\end{figure}

To better illustrate this behavior, Fig. \ref{Gra-ZiNeng} displays the curves of ${\rm Re}\Sigma^{(2)}(\bm{k},\varepsilon_{\bm{k}})$ and $\left|{\rm Im}\Sigma^{(2)}(\bm{k},\varepsilon_{\bm{k}})\right|$ along the high-symmetry line of $\rm \Gamma-M-K-\Gamma$.
The values of ${\rm Re}\Sigma^{(2)}(\bm{k},\varepsilon_{\bm{k}})$ and $\big|{\rm Im}\Sigma^{(2)}(\bm{k},\varepsilon_{\bm{k}})\big|$ are observed to be close to zero at the $\Gamma$ point, indicating the validity of the long-wavelength approximation.
As the wave vector $\bm{k}$ moves away from the $\Gamma$ point, Re$\Sigma^{(2)}(\bm{k},\varepsilon_{\bm{k}})$ exhibits a negative increase, consistent with the notion that magnons with larger momentum experience more significant renormalization.
When $\bm{k}$ approaches the first BZ boundary, Re$\Sigma^{(2)}(\bm{k},\varepsilon_{\bm{k}})$ exhibits pronounced oscillations.
Similarly, $\big|{\rm Im}\Sigma^{(2)}(\bm{k},\varepsilon_{\bm{k}})\big|$ also exhibits a gradual increase in the long-wavelength region and displays pronounced peak structures near the first BZ boundary.
Given that the magnon lifetime is proportional to the inverse of $\big|{\rm Im}\Sigma^{(2)}(\bm{k}, \varepsilon_{\bm{k}})\big|$, the behavior of $\big|{\rm Im}\Sigma^{(2)}(\bm{k}, \varepsilon_{\bm{k}})\big|$ indicates that, in the case of considering only magnon-magnon interactions, long wavelength magnons have relatively long lifetimes, while short wavelength magnons have limited lifetimes due to their strong scattering interactions with thermal magnons.
This phenomenon of long lifetime for long wavelength magnons and short lifetime for short wavelength magnons has been observed in many magnetic materials \cite{1-Bayrakci,1-Bayrakci2,1-Sun,1-Dietrich}.
Besides the magnon-magnon interactions, other interaction mechanisms, such as electron-magnon interactions \cite{1-Woolsey,1-Wesselinowa} and phonon-magnon interactions \cite{1-Sinha,1-Streib}, can also lead to a reduction in the lifetime of magnons.

In an effort to identify the underlying cause of the pronounced behaviors of $\Sigma^{(2)}(\bm{k},\varepsilon_{\bm{k}})$ near the first Brillouin zone (BZ) boundary, we calculate the scattering density of states \cite{1-Pershoguba}
\begin{eqnarray}
 W_{\bm k}=\frac\lambda N\sum_{\bm q}\delta(\varepsilon_{\bm k}+\varepsilon_{\bm 0}-\varepsilon_{\bm q}-\varepsilon_{\bm k-\bm q}),
\end{eqnarray}
Here $\lambda$ is an adjustable parameter used to ensure that the value of $W_{\bm{k}}$ is comparable to $\big|{\rm Im}\Sigma^{(2)}(\bm{k},\varepsilon_{\bm{k}})\big|$.
The $\bm{k}$ dependence of $W_{\bm{k}}$ (the blue dashed line) is illustrated in Fig. \ref{Gra-ZiNeng}.
The analysis reveals that the peaks of the $W_{\bm{k}}$ curve align with those of $\big|{\rm Im}\Sigma^{(2)}(\bm{k},\varepsilon_{\bm{k}})\big|$, indicating that the peak structure of $\big|{\rm Im}\Sigma^{(2)}(\bm{k},\varepsilon_{\bm{k}})\big|$ is primarily determined by the energy denominator in Eq. (\ref{Eq-ZiNeng-q}) and remains relatively unaffected by its vertex functions \cite{1-Pershoguba}.

\begin{figure}[t]
\centering
\includegraphics[scale=0.75]{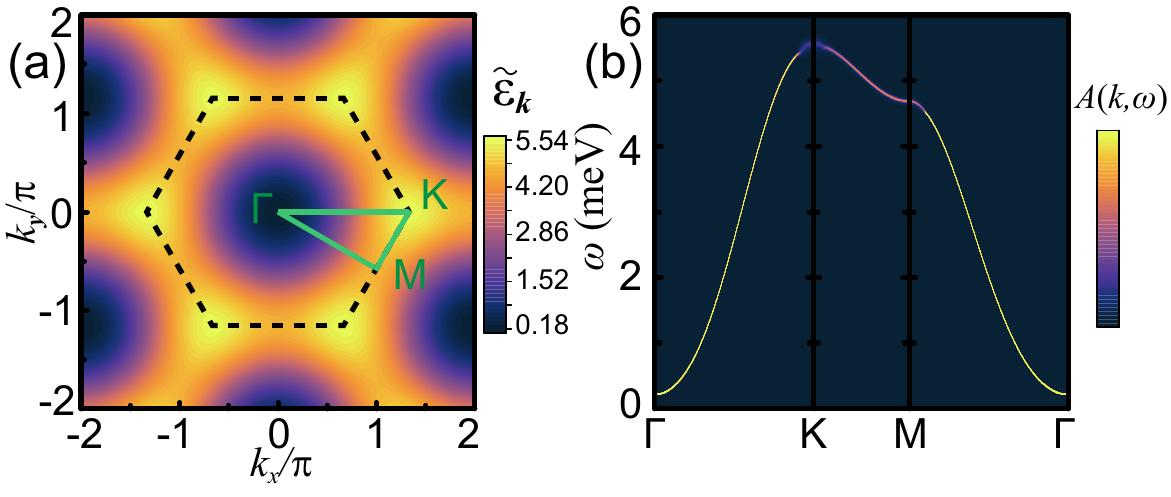}
\caption{(a) Magnon energy of MBT at $T=6K$ in the $\bm k$ space; the dashed hexagon indicates the BZ boundary of triangle lattice. (b) The map of magnon spectrum $A(\bm k,\omega)$ of MBT at $T=6K$ along the $\rm \Gamma-K-M-\Gamma$ line, where blue represents the lowest spectral intensity and red represents the highest spectral intensity.}\label{Gra-ek-Dispersion}
\end{figure}

\subsection{Renormalized magnon spectrum}

Taking into account the contributions from both $\Sigma^{(1)}(\bm{k})$ and ${\rm Re} \Sigma^{(2)}(\bm{k},\varepsilon_{\bm{k}})$ simultaneously, the dispersion $\tilde\varepsilon_{\bm{k}}$ in the $\bm{k}$ space at $T = 6K$ is depicted in Fig. \ref{Gra-ek-Dispersion}(a).
The highest magnon energy of $\tilde\varepsilon_{\bm{k}} = 5.54 meV$ occurs at the $\rm K$ point, while the energy saddle point is located at the $\rm M$ point.
Our findings indicate that the locations of the peak value and the saddle point of $\tilde\varepsilon_{\bm{k}}$ coincide with those of $\Sigma^{(1)}(\bm{k})$, demonstrating the significant impact of magnon-magnon interactions in the high-energy region \cite{1-Wei}.
This conclusion can be confirmed by analyzing the spectrum function of magnons, namely, the imaginary part of the magnon Green's function, which is defined as
\begin{eqnarray}
A(\bm k,\omega)=\frac{\left|{\rm Im}\Sigma^{(2)}(\bm k,\omega)\right|}{[\omega-\tilde\varepsilon_{\bm k}]^2+[{\rm Im}\Sigma^{(2)}(\bm k,\omega)]^2}.
\end{eqnarray}
Figure \ref{Gra-ek-Dispersion}(b) presents the map of $A(\bm{k},\omega)$ of MBT in the $\bm{k}-\omega$ space along the $\rm \Gamma-K-M-\Gamma$ line at $T=6K$.
The intensity of the magnon spectrum is low in the high-energy region, suggesting that magnons with high energy experience strong scattering interactions with thermal magnons.
Furthermore, the broadening of the dispersion line corresponds to the linewidth of the magnon spectrum.
Our calculations reveal that the broadening is pronounced near the boundary of the first BZ, with the most significant broadening observed around the $\rm K$ and $\rm M$ points.
This distinctive characteristic of the spectrum's linewidth awaits experimental verification.

\begin{figure}[htpb]
\centering
\includegraphics[scale=0.9]{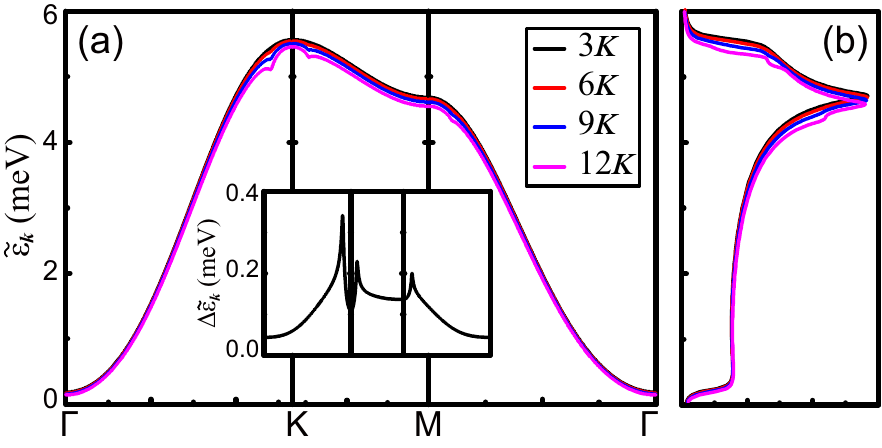}
\caption{(a) $\bm k$ dependence of $\tilde\varepsilon_{\bm k}$ along the $\rm \Gamma-K-M-\Gamma$ line of MBT at different temperatures. (b) The density of state of magnons of MBT at different temperatures.
The inset in (a) is the $\bm k$ dependence of $\Delta\tilde\varepsilon_{\bm k}$ of MBT along the $\rm \Gamma-K-M-\Gamma$ line.}\label{Gra-ek-dispersion-T}
\end{figure}

As the temperature increases, the renormalization of the magnon spectrum becomes more pronounced, accompanied by an enhancement in the frequency of magnon-magnon scattering.
In Fig. \ref{Gra-ek-dispersion-T}(a), the magnon energy dispersions along the high symmetry line $\rm \Gamma-K-M-\Gamma$ at different temperatures are plotted.
As the temperature increases, the magnon energy decreases overall due to the higher frequency of scatter events occurring between magnons.
A more pronounced effect from magnon-magnon interactions is observed around the boundary of the first BZ, indicating the $\bm{k}$ dependence of the renormalization of magnon spectra. 
The density of states (DOS) of magnons, $\int d{\bm k}\delta(\omega-\tilde\varepsilon_{\bm k})$, as a function of energy at different temperatures, is depicted in Fig. \ref{Gra-ek-dispersion-T}(b).
Three Van Hove singularities around $\omega\approx0.27eV$, $\omega\approx4.6meV$, and $\omega\approx5.4meV$ are observed, corresponding to the flat-band structure of the magnon spectrum at $\Gamma$, $\rm M$, and $\rm K$, respectively.
With an increase in temperature, all three peaks of the Van Hove singularities shift towards the lower energy region.

Furthermore, dip structures around the $\rm K$ and $\rm M$ points are observed on the $\tilde\varepsilon_{\bm{k}}$ curves at higher temperatures, as depicted in Fig. \ref{Gra-ek-dispersion-T}(a).
Correspondingly, several subpeak structures are observed around the two Van Hove singularities of higher energy in Fig. \ref{Gra-ek-dispersion-T}(b).
To visualize this structure more clearly, we define the difference between the magnon energy at low temperature and at high temperature:
\begin{equation}
\Delta\tilde\varepsilon_{\bm k}=\tilde\varepsilon_{\bm k}(3K)-\tilde\varepsilon_{\bm k}(12K).
\end{equation}
The $\bm{k}$ dependence of $\Delta\tilde\varepsilon_{\bm{k}}$ is illustrated in the inset of Fig. \ref{Gra-ek-dispersion-T}(a).
We observe that the peaks of $\Delta\tilde\varepsilon_{\bm{k}}$ coincide with the dips of ${\rm Re}\Sigma^{(2)}(\bm{k},\omega)$ at the same $\bm{k}$, providing evidence that the dip structures on the magnon spectrum are induced by the second-order self-energy. In some experiments, the dip structures of magnon spectrum near the boundaries of BZ have not been observed \cite{1-Bayrakci,1-Bayrakci2,1-Nikitin}. Therefore, further improvement is needed in the theoretical methods for studying the higher-order self-energy corrections of magnons.

\section{conclusion}

In this paper, we investigate the energy renormalization and lifetime of magnons in MnBi$_2$Te$_4$ caused by magnon-magnon interaction. By utilizing the perturbation theory of the many-body Green's function, the first-order self-energy [$\Sigma^{(1)}(\bm k)$] and the second-order self-energy [$\Sigma^{(2)}(\bm k,\varepsilon_{\bm k})$] of magnon are obtained. 
To avoid the false resulting from negative energy renormalization due to $\Sigma^{(2)}(\bm{k},\varepsilon_{\bm{k}})$, a long wavelength approximation of $\bm p\approx0$ is employed and a $T^2$ dependence of $\Sigma^{(2)}(\bm k,\varepsilon_{\bm k})$ is demonstrated analytically. 
Based on the numerical results, we find that the renormalizations from both $\Sigma^{(1)}(\bm k)$ and ${\rm Re}\Sigma^{(2)}(\bm k,\varepsilon_{\bm k})$ exhibit significant momentum and temperature dependence. Moreover, the energy renormalization near the first Brillouin zone (BZ) boundary is notably stronger compared to the center region, with the most prominent correction occurring near the $\rm K$ point. 
Furthermore, magnons near the center of the Brillouin zone possess almost infinite lifetime, whereas magnons near the boundary experience strong scattering. 
Finally, we demonstrate that the dip structures in the renormalized spectrum should be attributed to $\Sigma^{(2)}(\bm k,\varepsilon_{\bm k})$.

\section*{Acknowledgement}
The authors thank Professor Y. Song for the constructive suggestion and thank Professor H. Zhao, Dr. X. Tu, Dr. X. Zhu, and Dr. M. Zeng for the helpful discussion.
This work was supported by the NSFC (Grants No. 12188101, No. 11834006, No. 12147139, No. 12274184, and No. 12004170) and the China Postdoctoral Science Foundation (Grant No. 2021M691535).

\section*{Appendix: The details of the DFT calculation and the MC simulation}

To validate the reliability of $\bm k$-point mesh convergence, we perform first-principles calculations with different $\bm k$-point meshes from PBE-GGA calculations. The total energies and magnetic moments of the Mn ions calculated under different $\bm k$-point meshes are presented in Table \ref{APP-1}. 
\begin{table}[h]
\centering
\caption{The calculated total energy of MnBi$_2$Te$_4$ (Ry) and magnetic moments of Mn ions ($\mu_B$) from PBE-GGA calculations under different $\bm k$-point meshes.}\label{APP-1}
\begin{tabular}{| c | c | c |}
\hline
$\bm k$-mesh &  Total Energy (Ry) & Magnetic Moment ($\mu_B$)\\\hline
$4\times 4\times1$ & \quad-143016.74260346 \quad &  4.50486\\\hline
$8\times8\times1$ & \quad-143016.74363004 \quad   & 4.50492\\\hline
$12\times12\times1$ & \quad-143016.74371105 \quad & 4.50528\\\hline
$14\times14\times1$ & \quad-143016.74424296 \quad & 4.50525\\\hline
$16\times16\times1$ & \quad-143016.74422522 \quad & 4.50519\\\hline
\end{tabular}
\end{table}

We also supplement the Heisenberg exchange interactions under different $\bm k$-point meshes in Table \ref{APP-2}. According to the numerical results, we believe that a $16\times16\times 1$ $\bm k$-point mesh can provide a reasonably accurate description of the properties of the material. The nearest neighbor interaction is the dominant term, about 0.249 meV, which stabilizes the FM order. The exchange interactions between the next-nearest and the third-nearest neighbors are an order of magnitude smaller than those between the nearest neighbors, with values around -0.024meV and -0.01meV, respectively.
\begin{table}[h]
\centering
\caption{The calculated Heisenberg exchange interactions (meV) from PBE-GGA calculations under different ${\bm k}$-point mesh.}\label{APP-2}
\begin{tabular}{| c | c | c | c | c | c |}
\hline
 \quad   & $4\times4\times1$ & $8\times8\times1$ & $12\times12\times1$ & $14\times14\times1$ & $16\times16\times1$ \\ \hline
 $J_1$ & 0.26137 & 0.25131 &  0.24934 & 0.24924 & 0.24938\\\hline
 $J_2$ & -0.02308 & -0.02405 & -0.02423 & -0.02424 & -0.02428\\\hline
 $J_3$ & -0.00954 & -0.01008 & -0.01019 & -0.01019 & -0.01021\\\hline
\end{tabular}
\end{table}

Moreover, we perform the first-principles calculations using different exchange-correlation potentials including the standard PBE-GGA \cite{1-Perdew}, the Wu-Cohen generalized gradient approximation (WC-GGA) \cite{1-Wuwu}, and the modified Perdew-Burke-Ernzerhof generalized gradient approximation (PBEsol GGA) \cite{1-Perdew22}. The calculated values $J_1, J_2, J_3$ using these three exchange-correlation functionals of $16\times16\times1$ $\bm k$-point mesh are displayed in Table \ref{APP-3} and the results are close. Here, we choose the results obtained through the PBE-GGA.

\begin{table}[h]
\centering
\caption{The calculated Heisenberg interactions (meV) from different exchange-correlation potentials.}\label{APP-3}
\begin{tabular}{| c | c | c | c |}
\hline
\qquad\qquad\qquad & PBE-GGA & WC-GGA & PBEsol-GGA \\ \hline
 $J_1$ & 0.24938 & 0.29414 & 0.23157 \\\hline
 $J_2$ & -0.02428 &  -0.02461 & -0.02568\\\hline
 $J_3$ & -0.01021 & -0.00864 & -0.01181\\\hline
\end{tabular}
\end{table}

Here, we also provide some details concerning the estimation of $T_c$ through the MC simulations of the Metropolis algorithm for the Heisenberg model. The  $T_c$ values for MBT of different unit cells are obtained and presented in Table \ref{APP-4}. It can be observed that the values of $T_c$ show a slight decrease with the increase in unit cell numbers. Therefore, we are confident that the unit cell size of $28\times28\times1$ is adequately large and that the $T_c$ value of 15.9$K$ is reasonable.
\begin{table}[h]
\centering
\caption{The calculated $T_c$ ($K$) from MC simulation of different unit cells with periodic boundary conditions.}\label{APP-4}
\begin{tabular}{|  c  |  c  |  c  |  c  |}
\hline
 \quad  Unit Cells   \qquad   &  \quad  $T_c$ ($K$) \qquad &   \quad Unit Cells  \qquad   &  \quad   $T_c$ ($K$) \qquad  \\\hline
 $10\times10\times1$   &  16.7  &   $20\times 20\times1$   &  16.3 \\\hline
 $12\times 12\times1$  &  16.6  &   $24\times 24\times1$   &  16.1  \\\hline
 $16\times 16\times1$  &  16.4  &   $28\times 28\times1$   &  15.9  \\\hline
\end{tabular}
\end{table}

\end{document}